\documentclass[12pt]{article} 
\usepackage{amssymb,amsfonts,amsmath}
\usepackage{mathtext,cite}
\usepackage{enumerate,float}
\usepackage{graphicx}
\allowdisplaybreaks

\begin{document} 
\title{Amplification of coupled nonlinear oscillations of charged particle beam in crossed magnetic fields}
\author{{A.R.~Karimov$^{1,2}$, A.M.~Bulygin$^{2^*}$, P.A.~Murad$^{3}$}\\
$^1$Joint Institute for High Temperatures, \\
Russian Academy of Sciences,\\
Izhorskaya 13/19, Moscow 127412, Russia\\
$^2$National Research Nuclear University MEPhI, \\
Kashirskoye shosse 31, Moscow, 115409, Russia\\
$^3$Morningstar Applied Physics, LLC Vienna, VA  22182 USA\\
$^*$E-mail: {\underline{a.m.bulygin@gmail.com}}}

\date{ }
\newcounter{graf}
\maketitle{}

\begin{abstract}
A non-relativistic, charged-particle beam is placed into a crossed magnetic field. For such a system, the nonlinear electrostatic oscillations generation in the different degrees of the beam freedom may be triggered by the energy/momentum exchange between the beam’s particles and these external fields. The influence of oscillation dynamics of these fields and beam have been studied based on the cold-fluid hydrodynamic description.  As a result, the necessary conditions under resonant amplification of the beam’s natural oscillations are identified. Present results demonstrate that the beam density increases when the amplitude of radial and axial velocities increase. This process decreases the radius of the beam over the course of time. The technical application of the process applies in real accelerators such as a gyrotrons, FELs, and cyclotrons, where transverse size is limited by the size of the vacuum chamber. Thus redistribution of energy between the external field and the kinetic energy of the beam can effectively accelerate the beam by using an external magnetic field. These fields with both axial and radial directions use further this beam as an effective light source by identifying the resonance frequency to improve stability, focus particles, and wave propagation.
\end{abstract}

\maketitle
\section{Introduction}\suppressfloats 

The electric charge oscillating in a vacuum is the simplest source to generate electromagnetic waves. This effect lies within all modern devices from RF devices to FELs  \cite{HM,WD,RF_pulses,FEL,FEL_2,FEL_3}. It permits generating electromagnetic radiation with different intensity and orientation in the range of extremely high frequency to X-ray that is based upon charged particle beam oscillations. 

According to the "classical view" \cite{pan}, the generation of oscillations in the gyrotron  \cite{gir_l,automod_gir,gir_magn} is explained by a beam instability. The instability is based upon the rotation of electrons in the magnetic field that creates an electromagnetic wave with a resonance frequency leading to phase focusing particles and wave amplification. This idea technically uses an RF cavity which creates oscillations with a mode in the spectrum where the frequency is close to cyclotron frequency. Rotation of the particles inside the device is provided by the radial component of the magnetic field inside the RF cavity \cite{FEL_3}. The monoenergetic electron beam, which moves inside the undulator, is a light source in FELs \cite{CRIEB} and in cyclotrons \cite{synchr, Comp_synchr, X-men, KL, Ls}. Inside these devices,  electrons oscillate in an external periodic magnetic field. This creates electromagnetic waves with the frequency depending on electron energy (this is connected with the Doppler Effect when relativistic beams are being used) and parameters of a magnetic system. Moreover, the virtual cathode effect is being used to accelerate electrons in vircators \cite{virk_theory_1, virk_theory_2,radi_2,dinam_3}. The modulation of collector and reflected currents \cite{virk_1,dinam_3} and oscillations of turning points of particles that occurs during the generation process. In this case, the frequency of radiation is determined by electron beam frequency \cite{virk_2,radi_2}. It means, frequency depends only on the density of the flow. 

These devices have been used in technical and physical different ways for stimulating the electromagnetic generation. So it would be of interest to try to combine these methods of generation in merely one system. The present paper is an attempt to take a step in this direction. Here, we shall focus on the amplification of the natural coupled beam’s oscillations in the crossed magnetic fields of a specific type without considering  the internal  process of electromagnetic emission.  In essence, this means that we shall determine only the necessary conditions for generating electromagnetic radiation. 

\section{Physical formulation} \label{modeli}\suppressfloats

\begin{figure} 
\begin{center}
\includegraphics[width=10.cm]{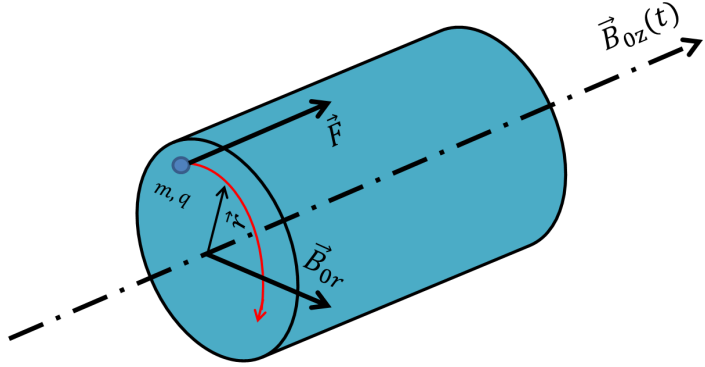}  
\end{center}   
\caption{Schematic of the charged particle beam acceleration region \label{Model_cool}}
\end{figure}

Let us consider a non-relativistic, rotating, cylindrical, charged particle beam moving in crossed external magnetic fields depicted in  Fig. \ref{Model_cool}, which was used before to accelerate plasma flows \cite{KM_17,KM_18,KM_19}. Here, the external magnetic field has a permanent radial component $B_{0{\rm r}}$ and the time-dependent axial component $B_{0{\rm z}}(t)$ generating the nonstationary azimuthal electric field which shall increase the beam rotation with respect to the z-axis. The interaction of such azimuthal flow with $B_{0{\rm r}}$ leads to a radial component of the Lorentz force that is directed opposite to the path of the Coulomb field of the beam. The rotating particle beam interacts simultaneously with the permanent radial component $B_{0{\rm r}}$ that leads to momentum transfer from the azimuthal degree of the freedom to the axial direction. 

It may be instructive to touch upon the congeniality of the schematic presented in Fig. \ref{Model_cool}, which is based upon routine electromagnetic generators.  Here, there is a magnetic field that will control oscillations inside a charged particle beam that works in a gyrotron and in FELs (here, a magnetic field will control oscillations inside a charged particle beam that works in a gyrotron and in FELs ). However, dynamic processes inside the beam \cite{dinam, dinam_2, dinam_3}
will determine the value of the frequency of inherent natural oscillations of the beam (so called  natural oscillations 
\cite{Ls}).

We suppose that these features will be able to lead to the generation of such oscillations as a function of the plasma density and velocity in the different macroscopic degrees of the freedom of the beam. There exists a certain relation between energy, frequency of natural oscillations of the beam and the value of the external magnetic field $B_{0{\rm z}}(t)$. The beam shall explicitly depend on degrees of the freedom of the oscillation. Such behavior is directly opposite to the usual dynamics of natural oscillations of the plasma density in the beams. We will study dynamics of an axis-symmetric ($\partial_{\varphi} \equiv 0 $) homogenous charged particle beam consisting with the same particles of $q$ charge and $m$ mass. In this framework, a cold-fluid hydrodynamic approach is considered to generate such coupled nonlinear oscillations by Lorentz force only described in Fig. \ref{Model_cool}.

In view of linearity of Maxwell equations, we can present the total electric and magnetic fields of the system as the sum of the internal and external components 
\begin{equation}
{\bf B}={\bf B}_{0} + {\bf B}^{*},
\label{Bfull}
\end{equation}
\begin{equation}
{\bf E}={\bf E}_{0} + {\bf E}^{*}.
\label{Efull}
\end{equation}
It is assumed there is an external spatially homogeneous magnetic field which depends only on time:
\begin{equation}
{\bf B}_{0}=B_{0{\rm r}}{\bf e}_{{\rm r}} + B_{0{\rm z}}(t){\bf e}_{{\rm z}},
\label{Bo}
\end{equation}
where $B_{0{\rm r}}=$const is the permanent radial magnetic field having a known value and $B_{0{\rm z}}(t)$ that is defined as some known time-dependent function.  From the induction equation for external fields:
\begin{equation}
\nabla \times {\mathbf{{E}_{0}}=-\frac{1}{c}\frac{\partial {\mathbf{B}_{0}}}{\partial t}},
\label{Farad_1}
\end{equation}
written for in integral form: 
\begin{equation}
\int\limits_\gamma {\bf E}_{0} d{\bf l}=-\frac{1}{c} \frac{\partial} {\partial {t}}\int\limits_{S\gamma} {\bf B}_{0}d{\bf S} ,
\label{intFarad}
\end{equation}
it follows that  for: 
\begin{equation}
\frac{\partial} {\partial {t}}\int\limits_{S\gamma} {\bf B}_{0}d{\bf S} \neq 0,
\label{magn_field}
\end{equation}
there appear conditions when an external magnetic field $B_{0}$ can create a vortex electric field $E_{0}$. To consider the axisymmetric of the flow presented in Fig. 1, we have:  
\begin{equation}
E_{0{\rm \varphi}}=-\frac{r}{2c}\frac{\partial B_{0{\rm z}}}{\partial t},
\label{E_0_phi_and_H}
\end{equation}
This means that the external electric field has only an azimuthal component. The intrinsic electric $\bf{E}^{*}$ and magnetic $\bf{B}^{*}$ fields are defined by the dynamic processes charged with a particle beam described by the standard cold-fluid model and Maxwell's equations:
\begin{subequations}
\label{cold_sys}
\begin{eqnarray}
&\dfrac{\partial n}{\partial t}+\nabla \cdot (n\mathbf{v})=0,
\label{Cold_n} \\
&\dfrac{\partial \mathbf{v}}{\partial t}+\left( \mathbf{v}\cdot \nabla \right) 
\mathbf{v}=\dfrac{q}{m}\left[ \mathbf{E}+\dfrac{1}{c}\mathbf{v}\times \mathbf{B%
}\right],
\label{Cold_v} \\
&\nabla \times \mathbf{{E}^{*}}=4\pi qn,
\label{Max_div_E} \\
& \nabla \times \mathbf{{E}^{*}}=-\dfrac{1}{c}\dfrac{\partial \mathbf{{B}^{*}}}{%
\partial t},
\label{Max_rot_E} \\
&\nabla \cdot \mathbf{{B}^{*}}=0,
\label{Max_div_B} \\
& \nabla \times \mathbf{{B}^{*}}=-\dfrac{4\pi }{c}qn{\bf v}+\dfrac{1}{c}\dfrac{%
\partial {\bf{E}^{*}}}{\partial t},
\label{Max_rot_B}
\end{eqnarray}
\end{subequations}
where $n$ and $\bf{v}$ are respectively the density and the velocity of the charged particle beam.

In order to write these equations as dimensionless variables, we used the initial density $n_0$, the initial radius of charged particle beam $R_0$, the inverse Langmuir is given by frequency $\omega^{-1} _{{\rm b}}={({4\pi n_0 q^{2}}/m)}^{-1/2}$, where $q$ and $m$ are a charge and a mass of the particle, as the natural scale of density, coordinates, and time. All velocities are normalized by the $\widetilde{V}=R_0\omega_{{\rm b}}$, the electric field is normalized to the ${\widetilde{E}}=(4\pi n_0 m \widetilde{V}^2)^{1/2}$, and  the magnetic field is normalized by the amplitude of external magnetic field $B_0$. As a result it one may rewrite Eq. (\ref{E_0_phi_and_H}) and \eqref{cold_sys} as
\begin{subequations}
\label{w_s_s}
\begin{eqnarray}
&E_{0{\rm \varphi}}=-\varkappa\dfrac{r}{2}\dfrac{\partial B_{0{\rm z}}}{\partial t},
\label{w_s_3} \\
&\dfrac{\partial n}{\partial t}+\nabla \cdot (n\mathbf{v})=0,
\label{w_s_4} \\
&\dfrac{\partial \mathbf{v}}{\partial t}+\left( \mathbf{{v}\cdot \nabla }%
\right) \mathbf{{v}}={\rm sgn}(q)[\mathbf{E}+\varkappa \mathbf{v}\times \mathbf{B}],
\label{w_s_5} \\
&\nabla \cdot \mathbf{E}^{*}={\rm sgn}(q)n,
\label{w_s_6} \\
&\nabla \times \mathbf{E}^{*}=-\varkappa \dfrac{\partial \mathbf{B}^{*}}{\partial t},
\label{w_s_7} \\
& \nabla \cdot \mathbf{B}^{*}=0,
\label{w_s_8} \\
& \varkappa \dfrac{\omega_{\rm L}^2}{\omega_{\rm b}^2}
\nabla \times \mathbf{B}^{*}={\rm sgn}(q)n\mathbf{v}+\dfrac{\partial \mathbf{E}^{*}}{\partial t},
\label{w_s_9} 
\end{eqnarray}
\end{subequations}
where $\varkappa$ is the dimensionless variable:
\begin{equation}
\varkappa = \omega _{{\rm c}}/\omega _{{\rm b}},
\label{cappa}
\end{equation}
here $\omega _{{\rm c}}=|q|B_0/mc$ and $\omega _{{\rm L}} = c/R_0$ are the frequency of the cyclotron resonance and characteristic frequency, respectively. Dimensionless value $\varkappa$ determines the different physical scripta for the system. It should be noted that in the real electrophysical devices will have $\varkappa \ll 1$, while the case $\varkappa \sim 1$ relates to the astrophysical phenomena with giant magnetic fields (e.g., for stars, black holes, and neutron stars in a binary pulsar).   

This nonlinear system of partial differential equations (PDE) \eqref{cold_sys} describes the dynamics of the cold non-charged particle beam moving in the external crossed magnetic fields in the form (\ref{Bo}). One could study different types of natural oscillations that propels in the external fields depending on the initial data for this system. 

We set the upper limit for $v_{r}(t)$, $v_{\varphi}(t)$ and $v_{z}(t)$ velocities on the order $\sim 0.1c$ because of we have to restrict on the non-relativistic case.

\section{A generalized model of Brillouin beam}

We define the set of the additional simplifications allow us to transfer the nonlinear system of partial differential equations to the system of ODEs for simplicity of our study. In particular, let the velocity components of the beam represent as:
\begin{equation}
{\bf{{v}}}=rA(t){\bf{{e}_{{\rm r}}}}+rC(t){\bf{{e}_{{\rm \varphi }}}}+rD(t){\bf{{e}_{{\rm z}}}},
\label{Sol_v_form}
\end{equation}
where $A(t)$, $C(t)$ and $D(t)$ are associated with the radial, axial, and azimuthal velocity components. Let us set density as:
\begin{equation}
n=n(t).
\label{Sol_n_form}
\end{equation}
Let us define the intrinsic electric field of the system as:
\begin{equation}
{\bf{E}^{*}}=r\varepsilon _{{\rm r}}(t){\bf{e}_{{\rm r}}}+r\varepsilon _{\varphi }(t){\bf{e}%
_{{\rm \varphi}}},
\label{E_in_pri_case}
\end{equation}
where $\varepsilon _{r}(t)$ and $\varepsilon _{\varphi}(t)$ are some unknown functions, which describes changes of radial and azimuthal components of the intrinsic electric field of the beam in time. We may not consider the axial component of the electric field in relation (\ref{E_in_pri_case}) if we assume:
\begin{equation}
\left|\frac{\partial E_{\rm z}/\partial z}{\partial E_{\rm r}/\partial r}\right| \ll 1.
\label{z_r_case}
\end{equation}
This means by physical sense that the kinetic energy of the beam is higher than the potential self-energy despite where the beam is considered as non-relativistic. Also, it should be noted that the model worked out is the Brillouin model of the beam  \cite{br}, which is usually used in the dynamics of such a beam \cite{Dv}-\cite{fsh2}. However, with nonstationary azimuthal electric field with a constant radial magnetic field, this provides a redistribution of momentum from azimuthal to axial component. It is also worth noting that similar ansatz has been used by Stanyukovich \cite{stn} for  studying  astrophysical  oscillations, and Amiranashvili  et  al. \cite{ays} and  Dubin \cite{dhed} for  studying  nonlinear oscillations  in  non-neutral  oblate  and  regular  spheroidal plasmas.

Substituting (\ref{E_in_pri_case}) into Faraday's law (\ref{w_s_7}) we get relations for components of intrinsic magnetic field
\begin{equation}
\frac{\partial B_{{\rm r}}}{\partial t}=0, \quad \frac{\partial B_{{\rm \varphi }}}{\partial t}%
=0, \quad \frac{\partial B_{{\rm z}}}{\partial t}=-\frac{2}{\varkappa}\varepsilon _{{\rm \varphi }},
\label{B_in_pri_case}
\end{equation}
This means the intrinsic magnetic field has only axial component under conditions $B_{r}(0)=0$ and $B_{\varphi}(0)=0$:
\begin{equation}
{\bf{B^{*}}}=B_{{\rm z}}(t){\bf{e}_{{\rm z}}},
\label{B_homog}
\end{equation}
and $\nabla \times \bf{{B}^{*}}\equiv 0$ (where the relevant discussion was presented in \cite{dinam, KM_17}). Dynamics of $B_{z}(t)$ is defined by the third relation in (\ref{B_in_pri_case}), i.e. depending on the sign of azimuthal  electric field where one can both increase and decrease of the total axial magnetic field. Moreover, we can rewrite (\ref{w_s_9}) as:
\begin{equation}
\frac{\partial \mathbf{{E}^{*}}}{\partial t}=-{\rm sgn}(q)n\mathbf{{v}.}
\label{E_con_nV}
\end{equation} 
It should be noted that in this case we will not use the relation (\ref{w_s_6}) in a further analysis because this relation is a linear superposition of relations (\ref{w_s_4}) and (\ref{E_con_nV}). 

By inserting (\ref{Sol_v_form}\mbox{--}\ref{E_in_pri_case}), (\ref{B_homog}) and inserting into \eqref{w_s_s}, (\ref{E_con_nV}) and using the expression (\ref{w_s_3}) yields:
\begin{subequations}
\label{End_eq_sys}
\begin{eqnarray}
&\dfrac{\partial A}{\partial t}+\left(A^{2}-C^{2}\right)={\rm sgn}(q)\left[\varepsilon%
_{r}+\varkappa C\left(B_{0{\rm z}}+B_{{\rm z}}\right)\right], \label{End_eq_1} \\
&\dfrac{\partial C}{\partial t}+2AC =  -\dfrac{\varkappa}{2}\dfrac{{\rm d}B_{0{\rm z}}}{{\rm d}t}+%
{\rm sgn}(q)\left[\varepsilon _{\varphi }+\varkappa \left(B_{0{\rm r}}D-\left(B_{0{\rm z}}+B_{{\rm z}}\right)A\right)\right], \label{End_eq_2} \\
&\dfrac{\partial D}{\partial t}+AD= -{\rm sgn}(q) \varkappa B_{0{\rm r}}C, \label{End_eq_3} \\
&\dfrac{{\rm d} n}{{\rm d} t}+2nA = 0, \label{Sol_n_res} \\
&\dfrac{{\rm d}B_{{\rm z}}}{{\rm d}t}= -\dfrac{2}{\varkappa}\varepsilon _{\varphi }, \label{End_eq_4} \\
& \dfrac{{\rm d}\varepsilon _{{\rm r}}}{{\rm d}t}= -{\rm sgn}(q)nA, \label{End_eq_5} \\
& \dfrac{{\rm d}\varepsilon _{{\rm \varphi }}}{{\rm d}t}= -{\rm sgn}(q)nC. \label{End_eq_6} 
\end{eqnarray}
\end{subequations}
This system of nonlinear ODEs describes the nonlinear processes in a charged particle beam occurring in crossed magnetic fields. However, it should be noted again that $B_{0{\rm z}}(t)$ is some known function, but its form will be defined afterward.
\section{Integral characteristics}

The approach for such an idea  (see Fig. \ref{Model_cool}) is  to propel the nonlinear electrostatic natural oscillations into the beam that requires spatial localization into space. This means that the transverse size of the beam doesn't need to be more than the transverse size of the beam pipe of the facility. So we should establish the moving boundary $R(t)$ for the beam, which separates the beam particles from the vacuum. Moreover, this spatial characteristic will be used to define such integral characteristics as the total momentum and total kinetic energy of the beam that is of interest from the experimental point of view.  

To find $R(t)$, we define the total number of particles as:
\begin{equation}
N=\int_{0}^{2 \pi} \int_{0}^{R} n(t) r d r d \varphi.
\label{Int_char_1}
\end{equation}
For spatial homogeneous density (\ref{Sol_n_form}) from (\ref{Int_char_1}) we have:
\begin{equation}
N=\pi R^2n.
\label{Int_char_2}
\end{equation}
This relation represents the law of particle conservation since there is no particles lost or particles-creation in the volume with the moving boundary $R(t)$ of the beam separating the beam from a vacuum. For simplicity let initial conditions be set as: $R(t=0)=n(t = 0)=1$. But one should remember that this may limit options. We will discuss this problem in future efforts. Then from (\ref{Int_char_2}), we get:
\begin{equation}
R(t)=\frac{1}{\sqrt{n(t)}}. 
\label{Int_char_3}
\end{equation}
Now one can calculate the total kinetic energy for the charged particle beam of the volume $R(t)$ by analogy with relation (\ref{Int_char_1}):
\begin{equation}
K(t)=\int_{0}^{2 \pi} \int_{0}^{R} \frac{v_{{\rm r}}^{2}+v_{{\rm \varphi}}^{2}+v_{{\rm z}}^{2}}{2} n(t) r \mathrm{d} r \mathrm{d} \varphi.
\label{Int_char_4}
\end{equation}
Inserting (\ref{Sol_v_form}), (\ref{Sol_n_form}) into (\ref{Int_char_4}) and using the expression (\ref{Int_char_3}) yields:
\begin{equation}
K(t)=\frac{\pi}{4 n(t)}\left({A(t)}^{2}+{B(t)}^{2}+{C(t)}^{2}\right).
\label{Int_char_5}
\end{equation}
\section{Dynamics of the coupled nonlinear oscillations} \suppressfloats   
As previously mentioned, the coupled azimuthal, radial and axial nonlinear oscillations can increase the external fields in the system shown in Fig. \ref{Model_cool}. Let us consider the model (\ref{Sol_v_form}\mbox{--}\ref{E_in_pri_case}), \eqref{End_eq_sys} where an axial component of the external magnetic field is changed by a harmonic law:
$$B_{{\rm 0z}}=B_{{\rm 0z}}{\rm{\sin}}(ht),$$ 
where $B_{{\rm 0z}}$ is the amplitude of the external magnetic field and $h$ is a dimensionless characteristic frequency. Here we get the process with a constant radial component of magnetic field directed outside of the beam. It means we will use the relation $B_{{\rm 0r}}=1$ for all further numerical solutions. A question immediately arises: how will changing oscillation characteristics affect where the natural oscillations influence on each other? This effect propels very strongly, especially near resonance frequency when the frequency of natural oscillations $\omega_{{\rm b}}$ is close to the frequency of the external field.  
\begin{figure} [htp!]
\begin{flushright}
\includegraphics[width=158.mm]{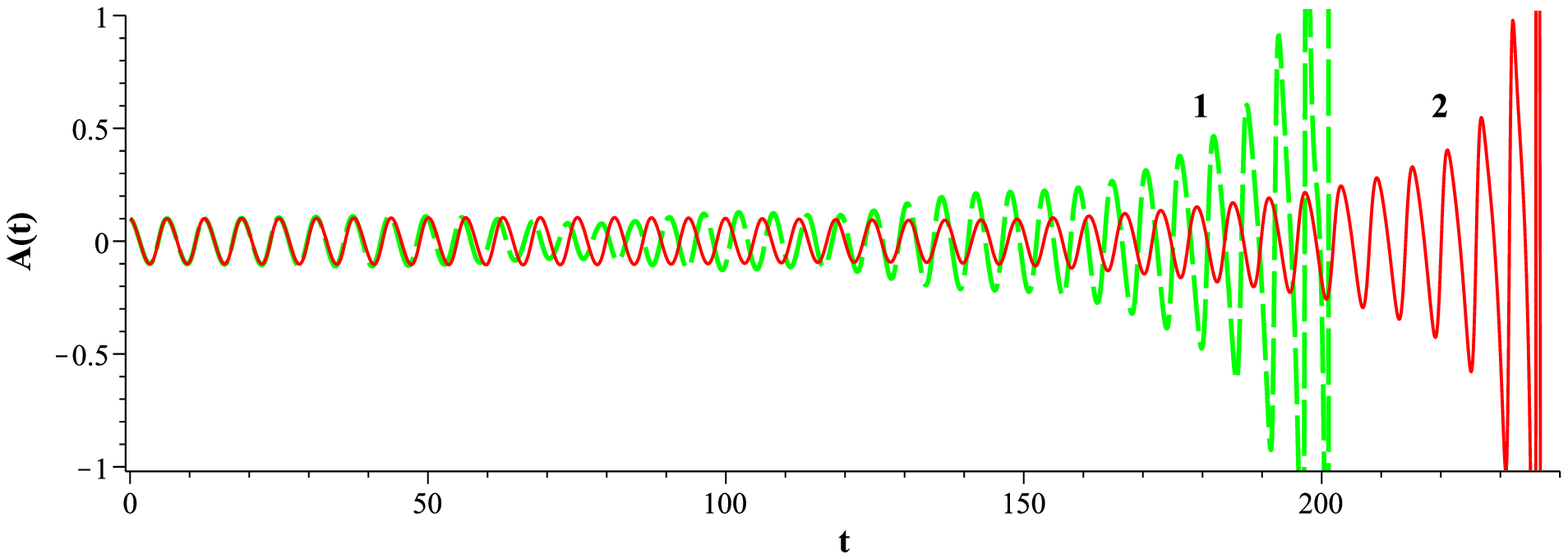}
\end{flushright}
\vspace*{-7mm} 
\begin{center}(a)\end{center}
\vspace*{-9mm} 

\begin{flushright}
\includegraphics[width=16.cm]{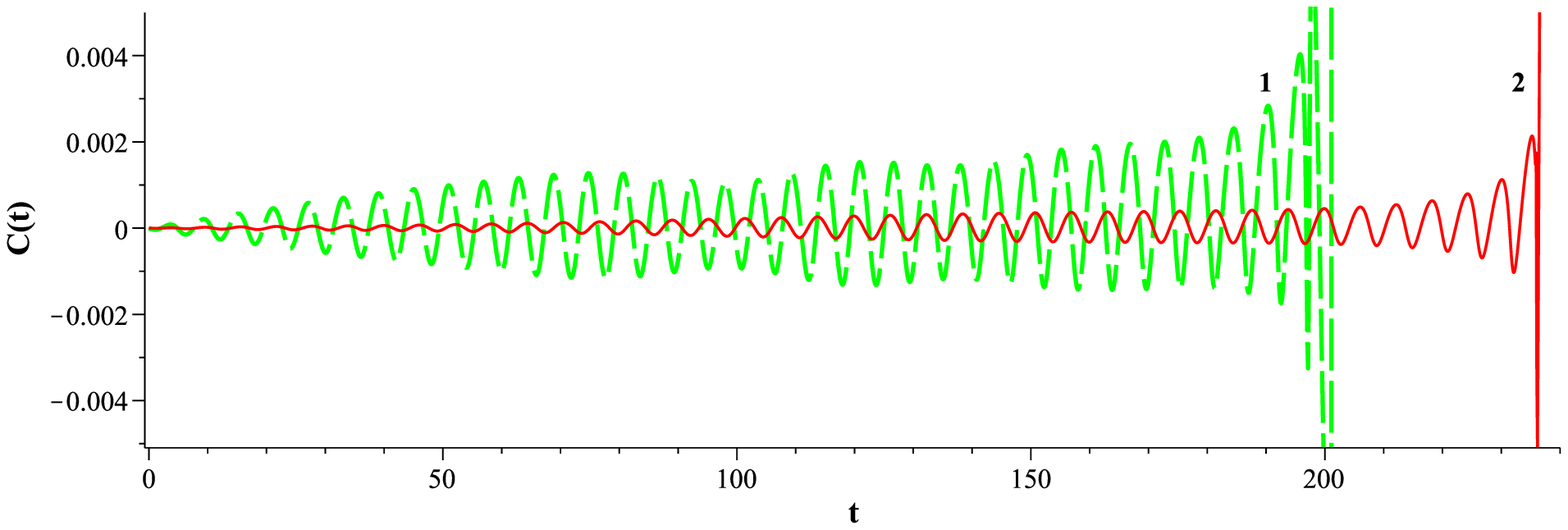} 
\end{flushright}
\vspace*{-7mm} 
\begin{center}(b)\end{center}
\vspace*{-9mm} 

\begin{flushright}
\includegraphics[width=157.mm]{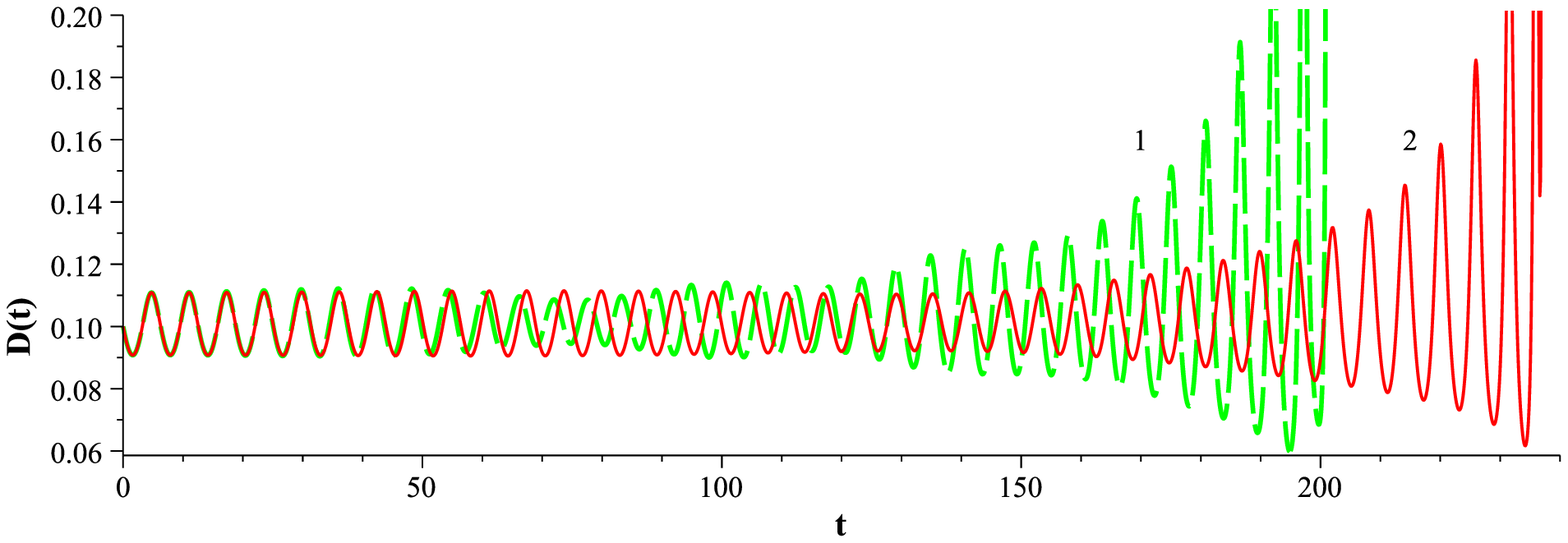} 
\end{flushright}
\vspace*{-7mm} 
\begin{center}(c)\end{center}
\vspace*{-5mm} 

\caption{The time dependence of (a) --- $A(t)$, (b)  --- $C(t)$ and (c) --- $D(t)$ for $A(0) = D(0) = 0.1$ and $C(0) = 0$. Curve 1 corresponds to $\varkappa = 10^{-4}$ (see Eq. \ref{cappa}); curve 2 corresponds to $\varkappa = 10^{-5}$. \label{Vel_comp}}
\end{figure}

Recall that $\varkappa$ is a small dimensionless parameter that relates to two frequencies. They are the frequency of the cyclotron resonance $\omega _{{\rm c}}=|q|B_0/mc$ and the characteristic frequency $\omega _{{\rm L}} = c/R_0$, respectively. This relation was mentioned before (see Eq. \ref{cappa}).   
\begin{figure}[ht]
 \begin{flushright}
\includegraphics[width=156.mm]{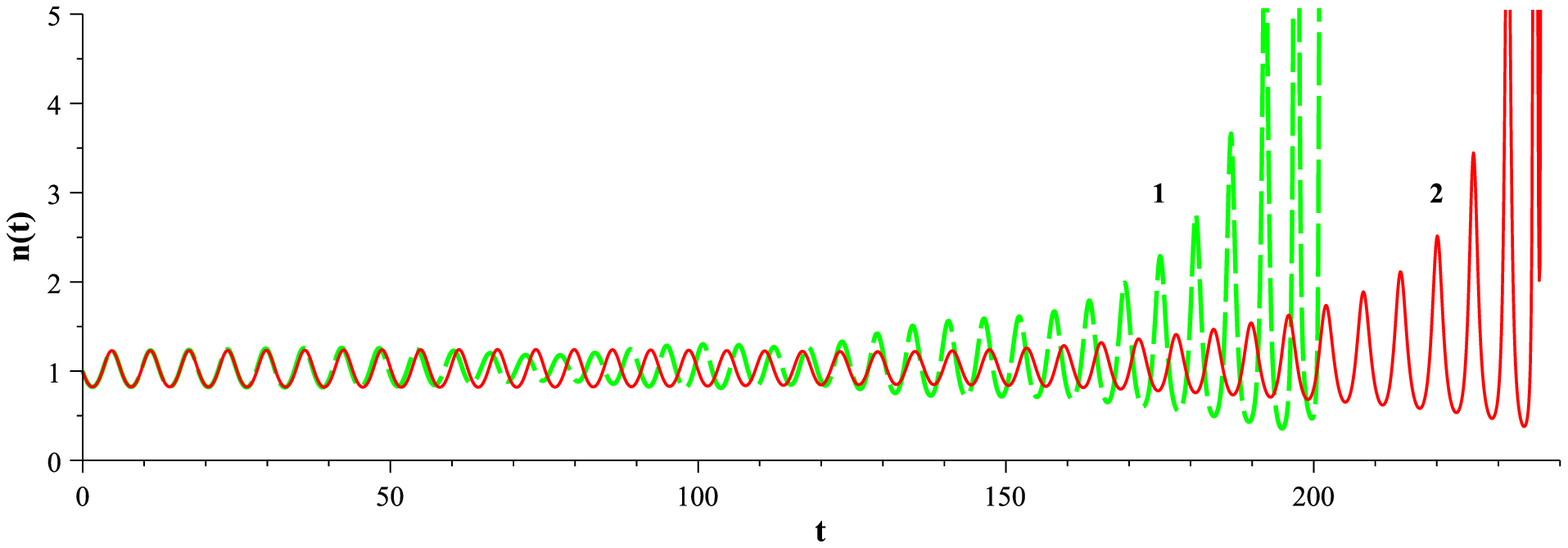} 
 \end{flushright}
\vspace*{-7mm} 
\begin{center}(a)\end{center}
\vspace*{-9mm} 

 \begin{flushright}
\includegraphics[width=158.mm]{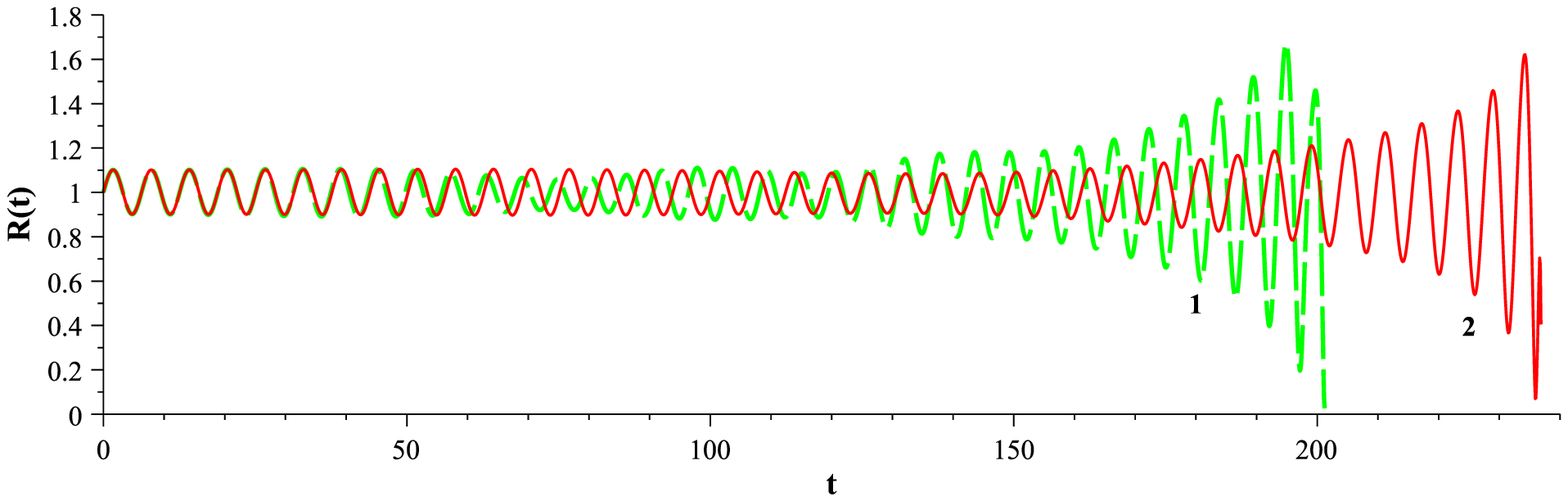} 
\end{flushright}
\vspace*{-7mm} 
\begin{center}(b)\end{center}
\vspace*{-4mm} 

\caption{The time dependence of (a) --- density $n(t)$ and (b)  --- radius $R(t)$ for $n(0) = R(0) = 1$. Curve 1 corresponds to $\varkappa = 10^{-4}$; curve 2 corresponds to $\varkappa = 10^{-5}$ (see Eq. \ref{cappa}). \label{beam_comp}}
\end{figure}
To demonstrate the possibility of such behavior, we analytically consider an electron beam (${\rm sgn}(q)=-1$) expanding in the external magnetic field with frequency $h \approx \omega_{{\rm b}}$ and $\varkappa \ll 1$. The external magnetic field plays a weak role in the dynamics of the process, but there is a generation of the intrinsic magnetic field (see Eq. (\ref{End_eq_4})). To understand the dependent results of the numerical solution on the $\varkappa$ value, it is convenient to rewrite relation (\ref{cappa}) as:
\begin{equation}
n_0=\frac{B_{0{\rm z}}^2}{4\pi m c^2 \varkappa^2},
\label{Par_p}
\end{equation}
which allows us to estimate ranges for $n_0$ and $B_{0{\rm z}}$ respectively. For example, if we let $\varkappa = 10^{-8}$ and $B_{0{\rm z}} = 1$~T, we will have the density $n_0 = 10^{8}$~cm$^{-3}$ and the value of the characteristic time is $\omega _{{\rm b}}^{-1} = 2 \times 10^{-9}$~c. The present paper is based on characteristics of real facilities and due to this we will consider $\varkappa$ values less than $10^{-4}$. Otherwise, the value of the density of the beam is extremely small.  For example, if we let $\varkappa = 10^{-3}$ and extremely high value of magnetic field $B_{0{\rm z}} = 30$~T (pulse mode), we will have only the density $n_0 = 10$~cm$^{-3}$. But if we consider these processes in stars where the magnetic field has potentially a limit close to $10^{11}$, one will be able to study processes for $\varkappa$ which vary near 1, but this is out of scope of this paper.
\begin{figure} [ht]
\begin{flushright}
\includegraphics[width=159.mm]{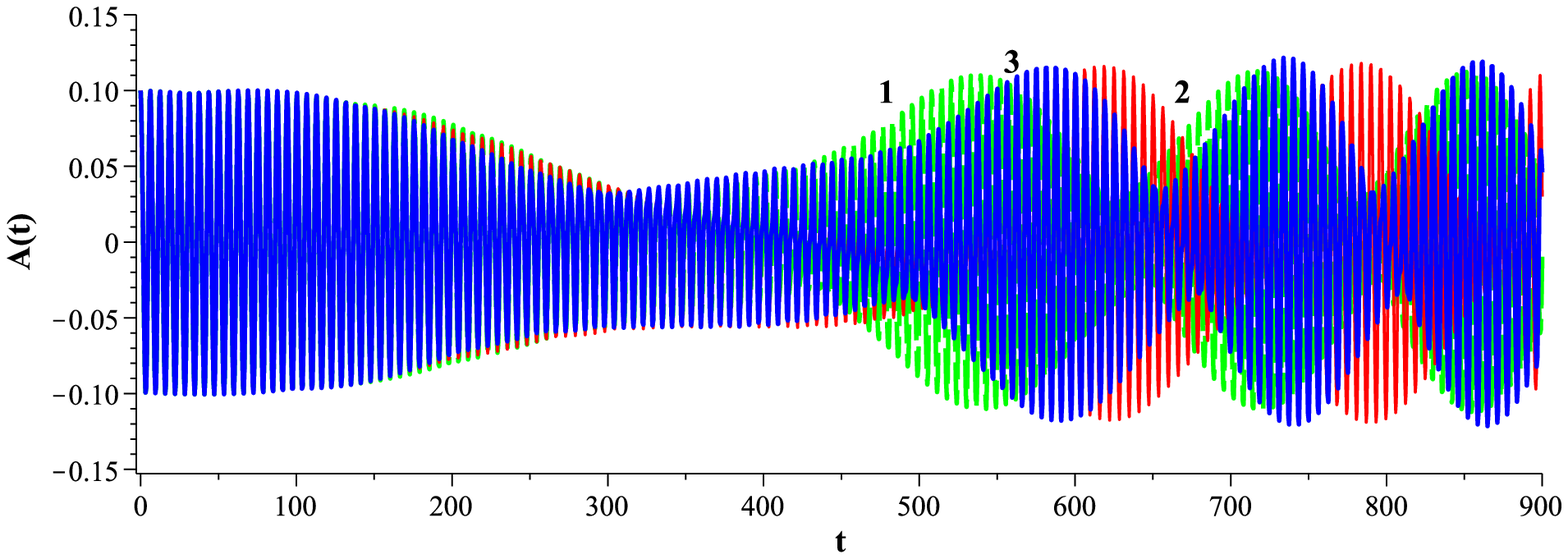}  
\end{flushright}
\vspace*{-5mm} 
\caption{The time dependence of $A(t)$ for $A(0) = 0.1$ and $\varkappa = 10^{-7}$ (see Eq. \ref{cappa}). Curve 1 corresponds to dimensionless characteristic frequency $h = 0.96$; curve 2 corresponds to $h = 0.97$; curve 3 corresponds to $h = 0.98$. \label{near_res}}
\end{figure}

\begin{figure} [ht!] 
\begin{flushright}
\includegraphics[width=159.mm]{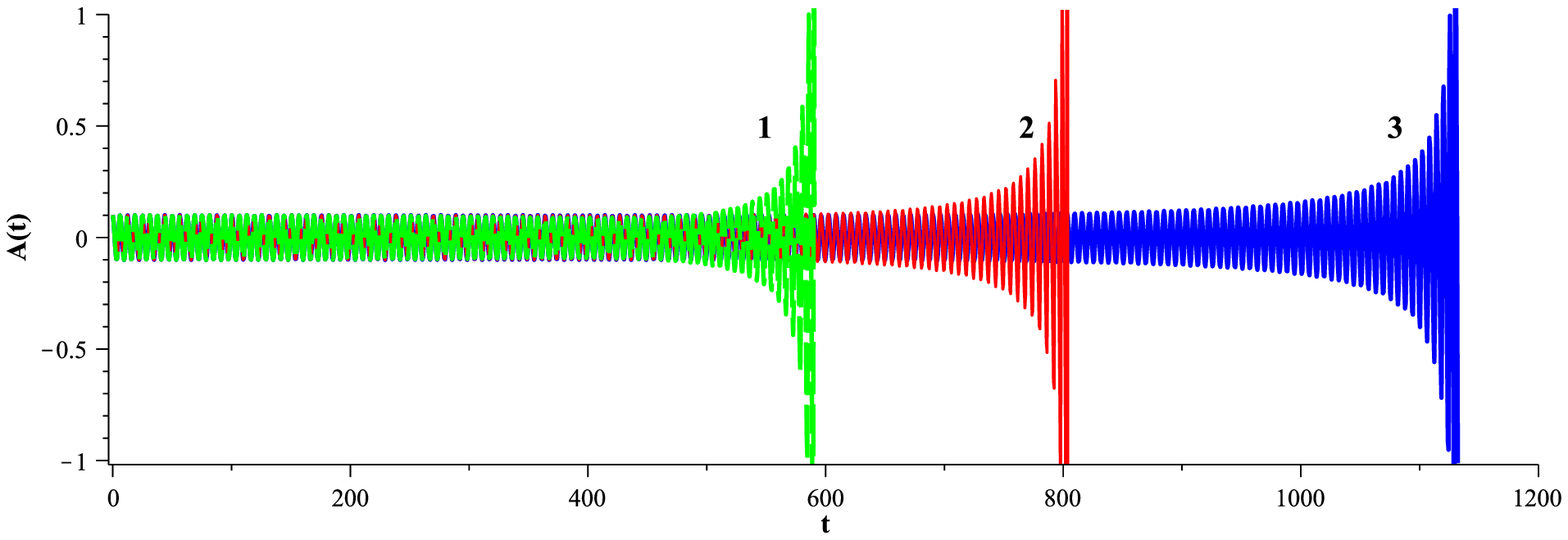}  
\end{flushright}
\vspace*{-5mm} 
\caption{The time dependence of $A(t)$ for $A(0) = 0.1$. Curve 1 corresponds to $\varkappa = 10^{-8}$; curve 2 corresponds to $\varkappa = 10^{-9}$; curve 3 corresponds to $\varkappa = 10^{-10}$ (see Eq. \ref{cappa}). \label{A(t)big_time}}
\end{figure}
We neglect the azimuthal velocity component in the electron beam at the initial moment of time for simplicity, i.e. we put $v_{{\rm \varphi}}(t=0) = 0$. Initial values of perturbed electric and magnetic fields were examined with $\varepsilon_{{\rm r}}(0)=\varepsilon_{{\rm \varphi}}(0)=B^{*}_{{\rm z}}(0)=0$. These conditions at the initial state in the quasineutral plasma shift, the density of electrons approximate for the density of ions $n_{e} \approx n_{i}$ but only one type of particles will leave the shift and will be accelerated. Initial values of radial and axial velocities were taken $v_{{\rm r}}(0)=v_{{\rm z}}(0)= 2\times10^{-3} c$ that approximately corresponds to 1 $eV$. That is the energy for electrons flying from the surface of the cathode or quasi-neutral plasma. Let us set value $\widetilde{V} = 0.02c$ for a more convenient graphic representation. In this case initial values of velocity components will equal $A(0)=D(0)=0.1$ and $C(0)=0$.

Results of the numerical solution with specified initial conditions show that generation of the coupled nonlinear oscillations occurs only in some specific points of the ($\varkappa, h$) plane for enough rarefied medium when $\varkappa > 10^{-8}$. Fig.\ref{Vel_comp} and Fig.\ref{beam_comp} represents examples of time evolution of the nonlinear oscillations for ($\varkappa=10^{-4}, h = 1.071$) and ($\varkappa  = 10^{-5}, h = 1.024$) cases. The coupled oscillations of radial, azimuthal and axial velocities are accompanied by a monotonic increase in their frequencies and amplitudes over the course of time. The growth of the amplitude of the velocity components provides large oscillations of the density $n(t)$ of the beam (see Fig. \ref{beam_comp}(a)) while the radius $R(t)$ is decreasing almost without limit (see Fig. \ref{beam_comp}(b)). This means that our model works correctly until density collapse when it appears in the beam.  In addition, functions $C(t)$, $D(t)$, $n(t)$  and $R(t)$ evolve in time similar to $A(t)$ so farther on we will represent $A(t)$ only as a radial velocity.

However, entirely opposite dynamics occurs for a lot of different values $h$ when $\varkappa > 10^{-8}$. One can see a typical behavior of the radial component of the field in Fig. \ref{near_res}, where initially ($t< 450$) occur, the decrease of oscillations in both radial and azimuthal components, but later ($t> 450$) there are formed wave trains (packets). Such behavior differs from the dependence in Fig. \ref{Vel_comp}, where one can see a constant increase in the amplitude. It seems this process is connected with momentum transfer from these oscillations to axial movement of the beam and it is going to saturate over time.

The generation of oscillations of the velocity and the density increasing with time does not depend on the $h$ value for the plasma flow density. This occurs with the increase of the amplitude of the velocity for several orders, as shown in previous plots. It means there are needed conditions for the generation of electromagnetic radiation. The decrease of $\varkappa$ leads to an increase in time of the singularity generation.   

It is worth to mention, that the decrease of $\varkappa$ leads to the exponential increase of time for all values in all cases (see Fig. \ref{A(t)big_time}). In addition, the increase of the amplitude of oscillations of different velocity components coincides with the increase of the frequency. This means that one can use the scheme as presented in Fig. \ref{Model_cool} for the generation of electromagnetic radiation. 

\section{Conclusion}   \suppressfloats    

In the present paper, we have considered the amplification of the coupled oscillations of finite amplitude for a charged-particle non-relativistic beam that is placed into crossed magnetic fields. Such a process is of interest for generating electromagnetic radiation; however, our consideration is restricted only by the simplest assumption of cold, non-relativistic, rotating charged particle beams without treating the intrinsic process of electromagnetic emission. Namely, we have defined only the necessary conditions for generating radiation.  

In the given schematic, the axial time-dependent component of the external magnetic field (used only the harmonic law) generates a nonstationary azimuthal vortex electric field. The azimuthal field provides momentum transfer from the field to the beam which leads to azimuthal acceleration of the charged particles. In addition, the permanent radial component of the external magnetic field leads to azimuthal momentum transfer to the axial direction that is coinciding parallel with the beam’s direction. This provides acceleration of the beam. Thus the increase of the kinetic energy of particles is caused by an interaction between the external field and charged particle beam. This leads to the increase of natural electrostatic oscillations as functions of the density and velocity of the beam in a specific relation between initial kinematic beam parameters. They are coupled both in the frequency and the amplitude of oscillations by relations (see Eq. \eqref{End_eq_sys}). 

One can see that such a process occurs when the frequency of the external field is close to the frequency of natural oscillations of the beam (see curves in Fig. \ref{Vel_comp}, \ref{beam_comp} and \ref{A(t)big_time}). There is an increase of the radial and axial components of the field for the present parameters of the beam and magnetic field in the current range of frequencies. As seen from these curves, the difference between the frequency of natural oscillations and the resonance frequency is less than 5\%. 

In addition, the present results demonstrate that the density of the beam increases over the course of time when the amplitude of radial and axial velocities increase. This process accompanies the decrease in the radius of the beam over the course of time (see Eq. (\ref{Int_char_3})). One can generate the process in real accelerators, where transverse size is limited by the size of the vacuum chamber. Thus the redistribution of energy between external field and the kinetic energy of the beam provides effective acceleration of the beam by using an external magnetic field with a special configuration in both axial and radial directions that may make it possible to use further this beam as an effective light source.

\end{document}